\documentclass[preprint,12pt]{elsarticle}

\usepackage{amssymb}
\usepackage{amsmath}
\usepackage{lineno}

\usepackage{standalone}
\usepackage{multicol}
\usepackage{multirow}
\usepackage{siunitx}
\usepackage{soul}
\usepackage{subcaption}

\usepackage{float}
\usepackage[section]{placeins}
\usepackage[hidelinks]{hyperref}
\usepackage{tikz}
\usetikzlibrary{positioning}
\usepackage[utf8]{inputenc}
\usepackage [style = english]{csquotes}
\MakeOuterQuote{"}
\biboptions{sort&compress}
\journal{Ënergy Conversion and Management}
\newcommand{\co}{CO\textsubscript{2} }
\newcommand{\Cone}{\SI{1.7}{\degreeCelsius} }
\newcommand{\Ctwo}{\SI{2.0}{\degreeCelsius} }
\begin{document}

\begin{frontmatter}



\title{Endogenous transformation of land transport in Europe for different climate targets}


\author[a,2]{Sina Kalweit} %% Author name
\ead{sk@mpe.au.dk}
\author[3]{Elisabeth Zeyen}
\author[2,4,5]{Marta Victoria}

\affiliation[a]{organization={Department of Mechanical and Production Engineering},%Department and Organization
            addressline={Katrinebjergvej 89}, 
            city={Aarhus},
            postcode={8200}, 
          %  state={},
            country={Denmark}}

\affiliation[2]{organization={Novo Nordisk Foundation CO2 Research Center},%Department and Organization
            addressline={ Gustav Wieds Vej 10}, 
            city={Aarhus},
            postcode={8000}, 
            country={Denmark}}

\affiliation[3]{organization={Department of Digital Transformation in Energy Systems, Faculty of Process Engineering, TU Berlin},%Department and Organization
            addressline={Einsteinufer 25 (TA 8)}, 
            city={Berlin},
            postcode={10587}, 
            country={Germany}}

\affiliation[4]{organization={Department of Mechanical and Production Engineering and iCLIMATE Interdisciplinary Centre for Climate Change},%Department and Organization
            city={Aarhus},
            postcode={8000}, 
            country={Denmark}}

\affiliation[5]{organization={Department of Wind and Energy Systems, Technical University of Denmark},%Department and Organization
            addressline={Elektrovej, 325}, 
            city={Lyngby},
            postcode={2800}, 
            country={Denmark}}
%% Abstract
\begin{abstract}
Road transport is responsible for about a quarter of Europe's greenhouse gas emissions, making its transformation a crucial part of Europe's overall decarbonization goals. Current European policies promote decarbonizing the transport sector and passenger car sales show an increased adoption of electric vehicles. Full electrification of land transport will significantly increase the average electricity demand but the use of smart charging and vehicle-to-grid could provide additional flexibility to balance wind and solar generation. In this study, we find cost-optimal transition pathways of the European land transport sector embedded in the sector-coupled open energy model PyPSA-Eur. We consider fossil-fueled, hydrogen-fueled, and electric cars using a 3-hour time resolution for a full year and covering 33 interconnected European countries. We analyze a transition path from 2025 to 2050 under different carbon budgets corresponding to a \SI{1.7}{\degreeCelsius} and \SI{2}{\degreeCelsius} temperature increase. Our results show that rapid electrification of road transport reduces the total system cost, even in the absence of climate targets. We see a clear preference for rapidly decommissioning internal combustion engine vehicles and using electric vehicles in all countries and under all carbon budgets. Allowing smart charging of electric vehicles decreases the total system cost by 1.6\% because it reduces the need to install stationary batteries by almost 40\%.
\end{abstract}


\begin{highlights}
\item Endogenous model for land transport in a European sector-coupled energy system
\item Myopic transition path to carbon-neutrality in 2050 
\item Road electrification in Europe is cost-optimal regardless of climate targets
\item Smart charging and V2G reduce costs by 1.6\% and stationary battery capacity by 61\%
\item Fuel cell cars are partially selected when they are 50\% cheaper or 21\% more efficient
\end{highlights}

%% Keywords

\end{frontmatter}

%\linenumbers

\section{Introduction}
A quarter of Europe's greenhouse gas emissions is due to road transport \citep{eea2023}. The European Commission has mandated to decarbonize the sector by allowing only zero-emission passenger cars and light commercial vehicles to be sold by 2035 \citep{fit455}. Electrifying the land transport poses extra requirements to the power system. First, by increasing the electricity demand. Second, it requires the deployment of charging stations, which may require reinforcing of the distribution grids. However, transport electrification can also provide additional flexibility to balance variable renewable generation. Electric vehicle (EV) batteries could be charged when there is an excess of renewable generation reducing curtailment. This is known as demand-side management (DSM) or V1G smart charging. Moreover, EV batteries could discharge into the grid to help supply the demand when needed. This is known as vehicle-to-grid (V2G).\\

Previous works have demonstrated a preference for electrification of land transport in decarbonized sector-coupled systems, though some consider hydrogen fuel cell electric vehicles (FCEVs) in a supplementary role \citep{Ruhnau2019, Brown2018}. It has also been shown that the optimal capacity for stationary batteries, when cost-optimizing a highly renewable power system, is significantly lower than the potential battery capacity of a fleet of EVs in Europe \citep{Brown2018, Victoria2019}. Therefore, it is plausible that only a share of EVs need to participate in DSM and V2G to help substantially balance renewable fluctuations. Moreover, the optimal electricity mix favors a higher share of solar photovoltaic (PV) generation when DSM of EV batteries is available due to the batteries' capacity to balance the PV generation daily cycle \citep{Victoria2019}. This synergy is particularly strong between rooftop PV systems and EV batteries, both connected on the distribution grid level \citep{Rahdan2024}. Despite the potentially large influence of EV batteries on the whole system, most previous analyses dealing with single countries \citep{Rinaldi2023} or a European sector-coupled energy system assume exogenous transformation of land transport \citep{Lund2021, Bogdanov2019, Wetzel2023, Brown2018, Neumann2023, Victoria2020, Victoria2022, Zeyen2023}. This assumption is especially limiting when a strict global \co target or budget is in place, as it restricts the efficient allocation of emissions across sectors. For some scenarios with stringent \co emissions limits, such exogenous assumptions drive significant demand for carbon-neutral liquid fuels to power the (remaining) internal combustion engine (ICE) vehicles \citep{Victoria2022, Zeyen2023}. Existing analyses that include endogenous transport modal shifts (e.g. from ICE vehicles to EVs or trains) are focused on one or a few countries and include low temporal resolution, which limits their capability to represent renewable balancing at different time scales properly \citep{Salvucci2019, Castelle2022}. The open model Euro-Calliope includes an endogenous transport modal shift to EVs but does not consider FCEVs because the authors assessed that current trends favor the use of EVs \citep{Pickering2022}. Nonetheless, recent EU regulations actively promote building a FCEV charging infrastructure for light-duty and heavy-duty vehicles along the TEN-T (Trans-European Transport Network) core network \citep{h2infra}.\\

To the best of our knowledge, no Europe-wide analysis including endogenous transformation of land transport exist. To address this gap, we model the transition of road transport with an endogenous decision regarding which powertrain (hydrogen, electric or internal combustion engine) is used in a sector-coupled European energy system. We investigate two different decarbonization pathways corresponding to a +\Cone and +\Ctwo warming. The energy system model PyPSA-Eur is used for a planning horizon of 2025-2050 with a 5-year investment cycle. This Europe-wide sector-coupled energy system model comprises electricity, heating, land transport, aviation, shipping, industry, and energy consumption in agriculture. We find synergy effects between sectors as well as differences between countries based on available renewable resources and demands. We also investigate the effect of allowing demand-side management, vehicle-to-grid, and different cost assumptions. 

\section{Methods and Materials}
\begin{figure}
    \centering
    \includegraphics[width=\textwidth]{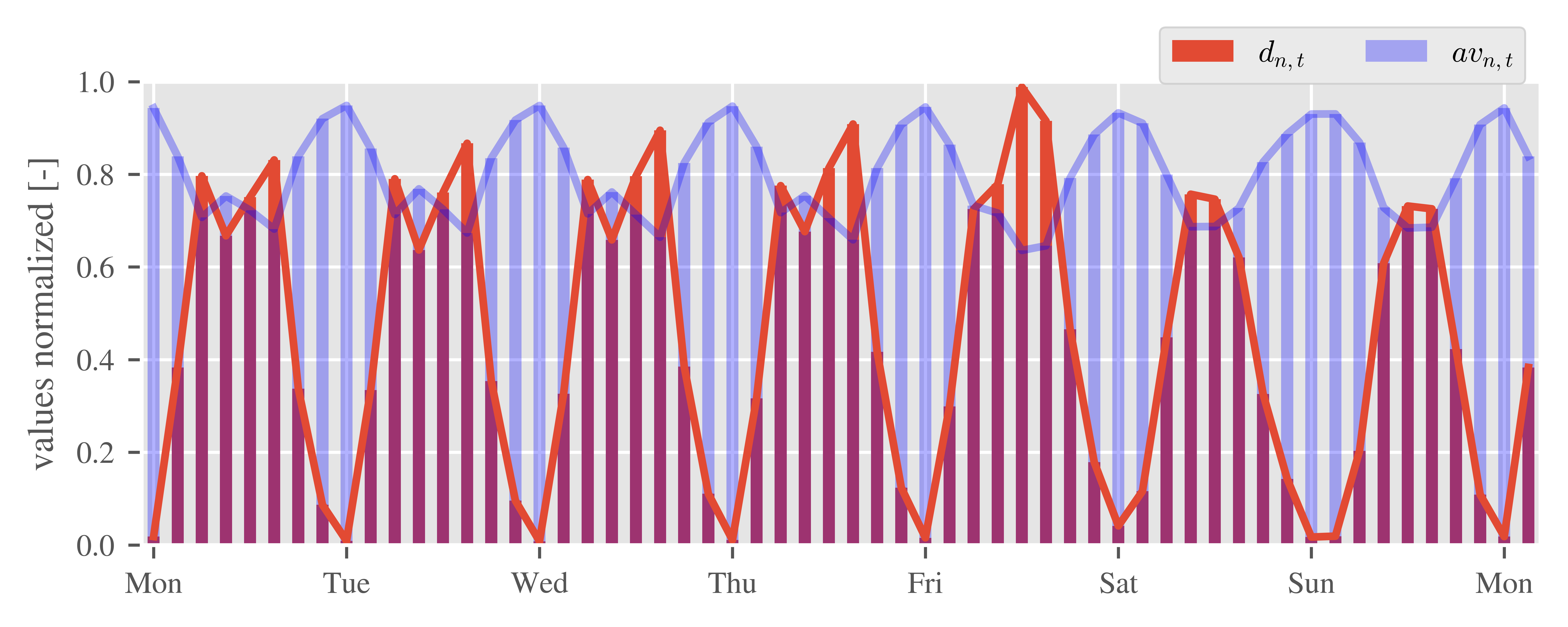}
    \caption{\textbf{Hourly profile of land transport demand and EV-grid availability:} The land transport demand $d_{n,t}$ is normalized to the maximum. The availability $av_{n,t}$ of EVs to interact with the grid shows the maximum capacity available in every time step (see Eq. \ref{eq:avail}) for charging and V2G use.
    }
    \label{fig:profile}
\end{figure}

PyPSA-Eur is an open energy system model of Europe. It optimizes the capacity and dispatch of generators, stores, and transmission lines, as well as the technologies required for providing demand in the electricity, heating, land transport, shipping, aviation, and industrial sectors (including feedstock), and energy demand in agriculture. Non-\co greenhouse gas emissions from agriculture are not included in the model, and it is assumed that emissions from this sector are offset by the land use, land use change, and forestry (LULUCF) sector. A detailed description of PyPSA-Eur can be found in Neumann et al. \cite{Neumann2023} and Victoria et al. \cite{Victoria2022}. The model clusters the existing ENTSO-E transmission grid to 37 nodes, one per asynchronous AC region and country. These nodes are connected by high-voltage AC or DC transmission lines, see Fig. \ref{fig:spatial}. In this work, the sector-coupled system is myopically optimized from 2025 to 2050 in five-year intervals, with each planning horizon having a 3-hourly temporal resolution for a full year and corresponding to a carbon budget. Myopically in this context means that the system is optimized sequentially and without knowledge of subsequent planning horizons. Europe’s carbon budget is derived by dividing the global budget equally per capita which results in a 6.43\% share for Europe. This share leads to a budget for Europe of \SI{57.9}{Gt} \co for \Ctwo and \SI{29}{Gt} \co for \Cone which is distributed over the years following an exponential decay and imposing carbon neutrality in 2050. See details on the carbon budget calculation in Note S1 in Victoria et al. \cite{Victoria2022}. The \co limit in every planning horizon depends on the overall carbon budget corresponding to the allowed future global temperature increase. 
 \\
\begin{figure}
    \centering
	 \includegraphics[width=\textwidth]{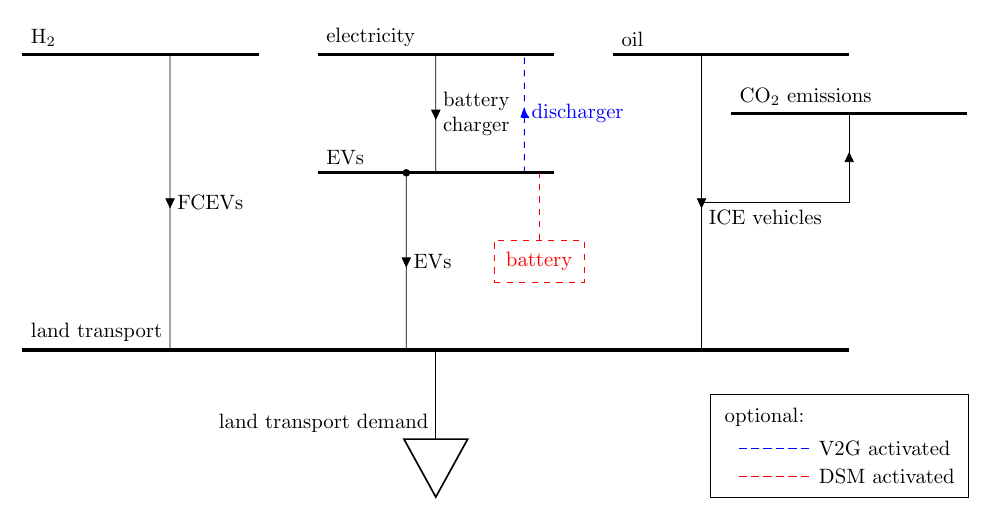}
    \caption{\textbf{Endogenous land transport implementation in PyPSA-Eur:} Each node in the model has a specific land transport demand (represented in kinetic energy) that must be supplied. A link in PyPSA allows converting energy with a certain efficiency, see \ref{Sup:3}. Three links (vertical lines, where arrows show the flow direction) represent FCEVs, EVs and ICE vehicles. ICE vehicles have two outputs, the land transport demand which they cover, and the \co which their use emits. The links representing the smart charging of the battery (demand-side management, DSM) and discharging (vehicle-to-grid, V2G) are only activated in some of the scenarios. For every link, its capacity, related to the number of cars, and dispatch, related to their usage, is optimized for each node.}
    \label{fig:links}
\end{figure}
In the following, we provide a brief summary of the model and a detailed description of the land transport sector. Unless otherwise noted, generation and conversion capacities are optimized by the model. Primary electricity generation technologies include onshore and offshore wind, solar photovoltaics, hydropower, biomass, coal, lignite, nuclear, gas and oil. The capacity of hydropower plants (reservoir and run-of-river generators) and pumped hydro storage are defined exogenously according to their historical capacities since we assume that the potential for expansion is limited in Europe. In this paper, we do not allow the expansion of the existing transmission network.
\begin{table}[ht]
\centering
\small
\begin{tabular}{l|l|l|l|l|l|l|l|l|l}
Type & Parameter & Unit & 2025 & 2030 & 2035 & 2040 & 2045 & 2050  \\\hline
\multirow{3}{*}{EV} & Capital costs & €/vehicle &  \num{28812} & \num{24624} & \num{24358} & \num{24092} & \num{23827} &\num{23561} \\
&Lifetime & years& 15& 15& 15 &15 &15 &15 \\
&Consumption &kWh/km& 0.2& 0.2& 0.2& 0.2& 0.2& 0.2\\\hline
\multirow{3}{*}{ICE}
&Capital costs &€/vehicle  &\num{24309} &\num{24999} &\num{25622} &\num{26167}& \num{26610}& \num{26880}\\
&Lifetime & years&  15&  15& 15 &15 &15 &15\\
&Consumption &kWh/km & 0.66& 0.66& 0.66& 0.66 &0.66& 0.66\\\hline
\multirow{3}{*}{FCEV}
&Capital costs &€/vehicle  &\num{43500} &\num{33226}& \num{30720}& \num{29440} &\num{28160} &\num{26880}\\
&Lifetime & years&  15& 15& 15 &15 &15 &15\\
&Consumption &kWh/km& 0.28 &0.28 &0.28 &0.28 &0.28 &0.28\\
\end{tabular}
\caption{\textbf{Vehicle parameters assumed in the model.} Capital costs and lifetime from Sterchele et al. \cite{Sterchele2020}. See \ref{Sup:1} for details on vehicle consumption. }
\label{tab:parameters}
\end{table}
In previous analyses \cite{Victoria2022, Neumann2023, Victoria2020, Zeyen2023}, the share of internal combustion engine (ICE) vehicles, electric vehicles (EVs), and hydrogen fuel cell electric vehicles (FCEVs) in the land transport sector was determined exogenously. Here, for the first time, we implement an endogenous optimization by the model depending on the costs and efficiencies of every type of vehicle. The transport demand can be supplied by ICE vehicles, FCEVs, or EVs, see Fig. \ref{fig:links}. The vehicles are aggregated per node and represented using a lumped demand profile of all vehicles within that region, Fig. \ref{fig:profile}. Given the challenges of inconsistent data across European countries and the computational burden of modeling multiple sectors with high detail, certain assumptions are necessary for modeling land transport alongside other sectors. For simplicity, the demand of the land transport sector comprising passenger vehicles, 2-wheelers, light commercial vehicles, heavy commercial vehicles, and non-electric rails is represented by passenger vehicles.\\

The land transport demand is based on the hourly-resolved weekly demand profile from the German Federal Highway Research Institute (BASt) \cite{bast}. Using the number of vehicles and final energy consumption for the total road and non-electric rail transport from the JRC-IDEES 2015 database \cite{JRC-IDEES}, the demand profile is scaled to the demand of each country. The assumptions for capital costs, lifetime and consumption are shown in Table \ref{tab:parameters}. Cost assumptions for EVs are uncertain and the values strongly depend on the source. While Sterchele et al. \cite{Sterchele2020} assumed a cost of \SI{28812}{EUR} for a standard EV in Europe in 2025, average costs in 2024 were around \SI{45000}{EUR} (2024) \cite{transportenvironmentWhatsWrong}. Moreover, worldwide vehicle capital costs differ drastically. While around 65\% of EVs sold in China reached price parity with ICE vehicles in 2023, according to the IEA's calculation, EVs in the USA remain more expansive than their conventional counterparts \cite{IEA2024}. This may have strong consequences if the imports of EVs in Europe changes in the future due to higher import taxes or restrictions and if production in Europe cannot keep up with the demand. In 2023, 1.4 million EVs were produced in the three highest European car production countries Germany, Spain and France while 1.5 million EVs were newly registered in the EU \cite{carproduction}. Differences in taxing of diesel, petrol and electricity prices may also impact competitiveness of vehicles. To deal with cost uncertainty, we implement a sensitivity analysis, see Fig. \ref{fig:bevavail}.\\
For 2025, we assume that ICE vehicles supply 96\% of the transport demand and EVs have a 4\% share, based on the assumption that the European share of 1.7\% of EVs and 1.3\% of plug-in hybrid EVs of 2023 will be slightly increased 2025; thereafter the model optimizes endogenously the ratios of the different vehicle types. The model can decommission vehicles before they reach the end of their lifetime, but the capital costs of those vehicles are still included in the total cost. For the transition path, ICE vehicles are uniformly distributed over the years 2011 to 2025, ensuring that the first 1/15 reaches the end of its lifetime in 2026, while the last 1/15 needs to be replaced after 2035. The already existing EVs are distributed over the years 2021 to 2025 to respect their low representation in earlier years. From 2030 on, the model optimizes the share of ICE vehicles, EVs or FCEVS per region endogenously based on their consumption costs and capital costs, see detailed description in Supplementary Note 3. The consumption costs depend on vehicle-specific consumption per km and regional and time-dependent hydrogen, electricity or oil prices. The hourly usage for every link represents all the vehicles driving in a region in every hour and it is required that each type of vehicles follow the demand curve (Fig. \ref{fig:profile}), preventing that one type of vehicle is only used for part of the year.\\
EVs have two flexibility options: DSM (demand-side management) and V2G (vehicle-to-grid). In this paper, a sensitivity analysis for the availability of V2G and DSM is performed. First, if DSM is activated, the battery can charge when it is cost-optimal for the whole system. Second, if V2G is activated, the EV battery can be discharged back into the grid. For both charging and discharging, an efficiency of 90\% is assumed. The capacity of an individual EV battery is assumed to be $E_{\text{EV\:battery\:capacity}}=$\SI{67}{kWh} \cite{Sterchele2020} and the DSM and V2G participation in our base scenarios is $p_{\text{DSM}}= 50\%$ of the available EVs per region. We assume that EVs not participating are charged directly after driving. Finally, the link "EVs" converts the electricity into kinetic energy to supply land transport demand.\\
To represent EV batteries availability, we assume the same constraint as previously described \cite{Brown2018}, where EV charging availability is dependent on land transport demand, see Fig. \ref{fig:profile} and if DSM is activated, the lumped EV battery needs to be at least 75\% charged at 6am, see \ref{Sup:3}. We impose three new constraints here:
First, the optimized charging capacity $F_{\text{EV charger}, n}$ is multiplied the DSM share $p_{\text{DSM}}$ and discharging capacity $F_{\text{V2G},n}$ of the lumped EVs in a node are set to be the same:
\begin{equation}
    F_{\text{EV charger}, n}\cdot p_{\text{DSM}} = F_{\text{V2G},n}\quad \forall n \label{eq:vsg}
\end{equation}\\
Second, the cost optimization determines the minimum number of EVs, together with the optimal number of FCEVs and ICE vehicles, that are needed to satisfy the land transport demand. However, we know that the current number of vehicles in Europe is far from the number required if they were optimally utilized. For example, in Europe vehicles are estimated to be, on average between 92\% and 97\% of the time parked \cite{parkingEVs,Bates2012}. To accommodate for this, we allow the battery power capacity of the EVs to be extended beyond the number of EVs strictly needed to satisfy the land transport demand. The ratio $F_{EV,n}/elec_{1vehicleEV}$ indicates the number of EVs strictly needed to supply the transport demand, which has been cost-optimally determined. We assume that the number of EVs that would exist in reality and can participate in DSM and V2G is 5 times higher. This is based on our assumption that only 20\% of vehicles are driving during peak demand in reality \cite{Brown2018}.\\
Hence, the lumped charging capacity for EVs battery is:
\begin{equation}
    \frac{F_{\text{EV charger},n}}{r_{\text{charge}}}  = 5\frac{F_{\text{EV},n}}{elec_{1 vehicle EV}}\quad \forall n \label{eq:BEV_charger}
\end{equation}
Third, the lumped charging capacity $F_{EV charger, n}$ and energy capacity $E_{EV battery, n}$ of the EV battery in every node $n$ is related through the unitary charging capacity $r_{charge}$ and energy capacity $E_{\text{EV\:battery\:capacity}}$.:
\begin{equation}
    \frac{E_{\text{EV battery},n}}{E_{\text{EV\:battery\:capacity}}} = \frac{F_{\text{EV charger},n}\cdot p_{\text{DSM}}}{r_{\text{charge}}}\quad \forall n \label{eq:dsm}
\end{equation}

\section{Results}
We start by presenting the results of our base scenarios, in which we allow DSM (demand side management) and V2G (vehicle-to-grid) for a carbon budget corresponding to \Cone and \Ctwo temperature increase. Section \ref{sec:1} describes the transition, system costs, EV charging and V2G use pattern. We continue in section \ref{sec:2} with a cost and consumption sensitivity analysis for EVs and FCEVs and finish in section \ref{sec:3} by showing the regional differences in the transformation for the base scenario and for a FCEV 50\% cost reduction case.\\
\subsection{Transformation of land transport for different carbon budgets.}
\label{sec:1}
\begin{figure}[htbp]
        \centering
        \includegraphics[width=\textwidth]{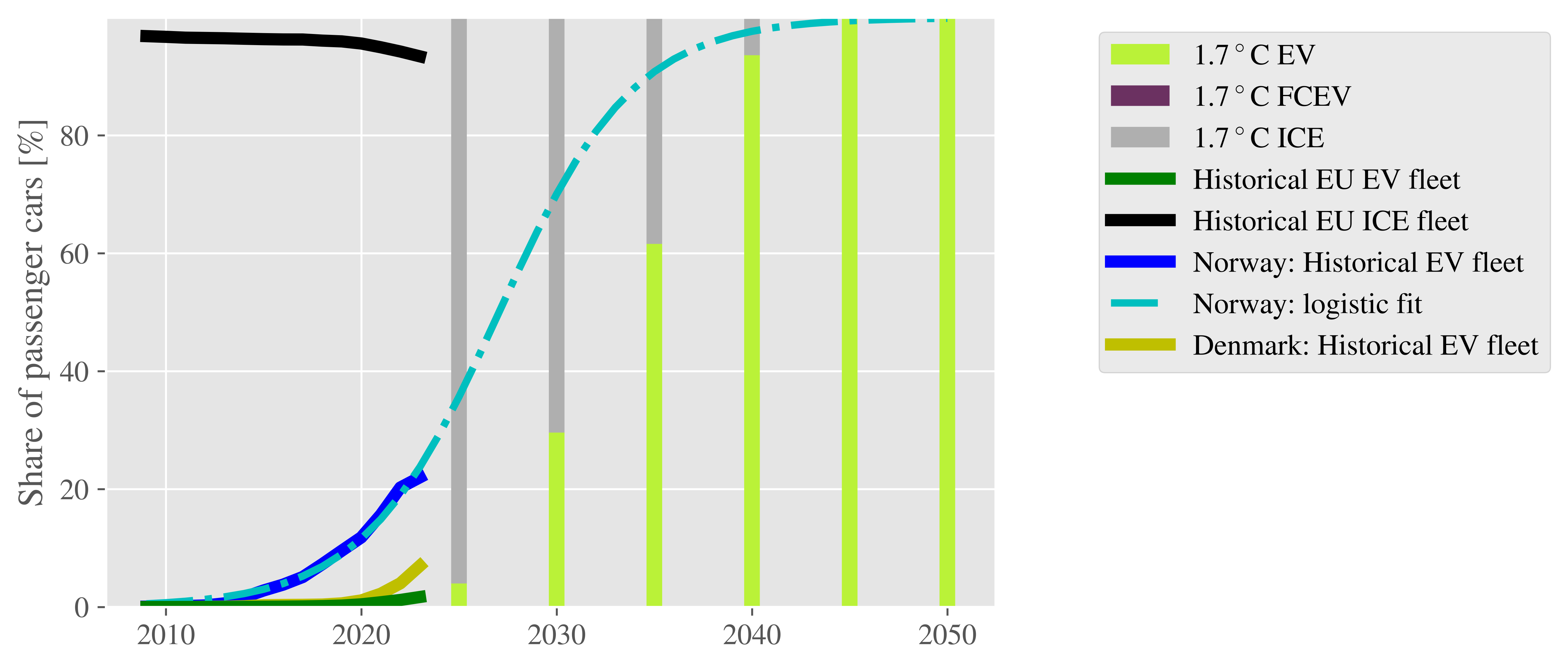}
\caption{\textbf{Fast transition to EV} Share of EVs and ICE vehicles in a cost-optimal transition for a \Cone carbon budget. The figure also shows the historical evolution of the European-wide EV fleet \cite{afo}, the EV fleet in Norway and Denmark\cite{afo}, and a logistic projection for the fleet in Norway.}
\label{fig:Evtransition}
\end{figure}
Allowing the model to choose the fuel type for vehicles from 2030 on, leads to an early substitution of the fossil-fueled fleet to EVs, both for the \Cone and the \Ctwo transition path. The \Cone transition path is shown in Fig. \ref{fig:Evtransition}, where around 30\% of the fleet in 2030 is electric, increasing to 94\% in 2040 until fossil-fueled vehicles are completely replaced in 2045. EVs are deployed to fill the gap left by ICE vehicles when the latter are retired. This also happens for the \Ctwo transition path and even when no \co constraint is imposed (Fig. \ref{fig:car15}). In comparison, the historical evolution of the European-wide fleet only reached 1.7\% in 2023 while Norway and Denmark have, with a share of 22\% and 7\%, respectively, the highest percentage of EVs in Europe \cite{afo}. Assuming a logistic growth, the transition to EVs in Norway would follow a similar transition speed than that required for the \Cone transition path.\\
\begin{figure}[htbp]
\begin{minipage}{0.33\textwidth}
        \caption*{\small{\Cone}}
\end{minipage}
\begin{minipage}{0.11\textwidth}
    \caption*{}
\end{minipage}
\begin{minipage}{0.45\textwidth}
        \caption*{\small{\Ctwo}}
\end{minipage}
\centering
\includegraphics[width=\textwidth]{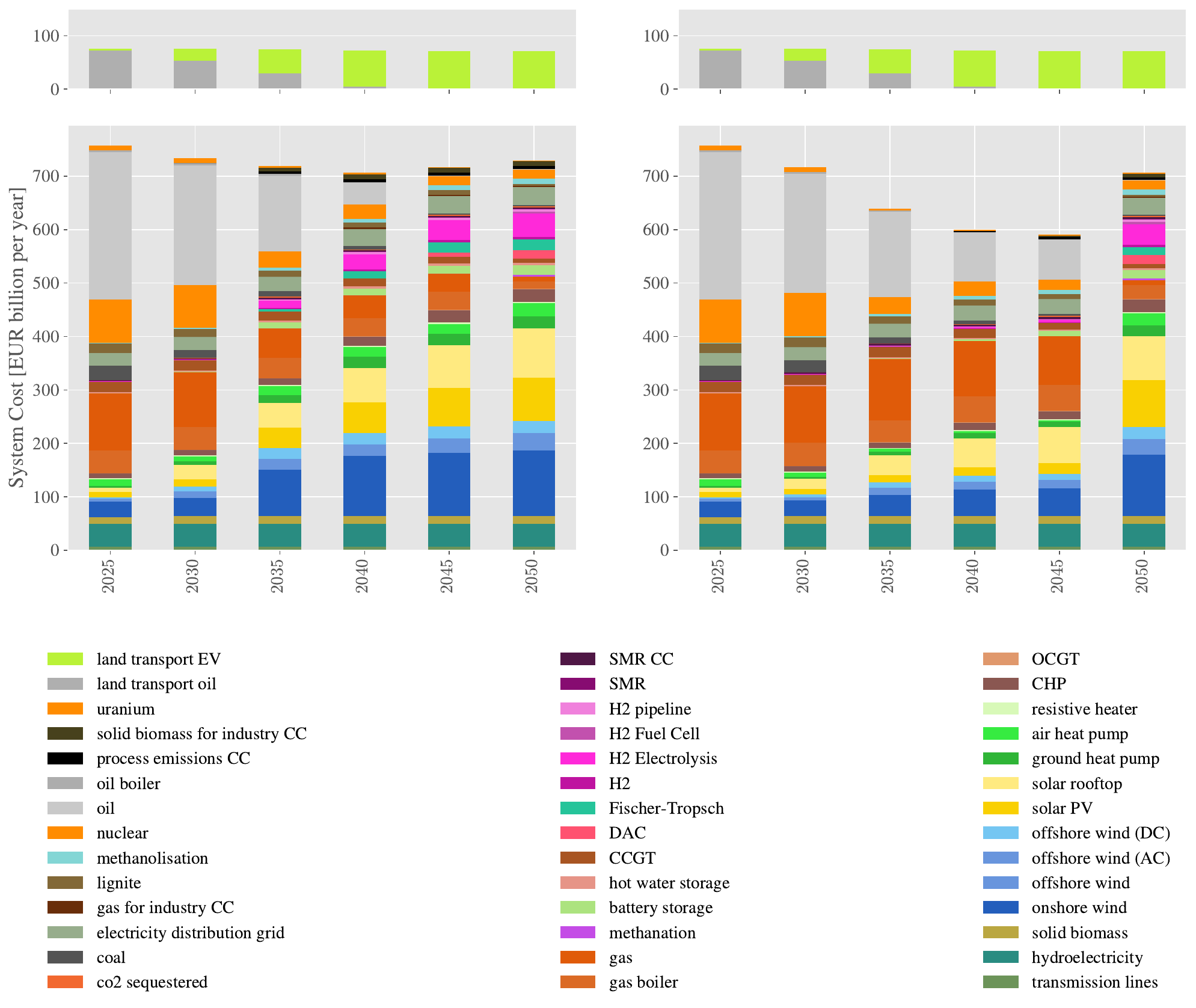}
\caption{\textbf{Total system costs:} Distribution of total system costs for a carbon budget corresponding to (left) a \Cone temperature and (right) a \Ctwo temperature increase. The land transport capital costs are shown separated in the upper row while fuel costs are included in the overall system costs shown at the bottom. In 2025, the fleet was exogenously set to contain 96\% of ICE vehicles and 4\% of EVs. The land transport capital costs include only the cost of vehicles strictly needed to supply the land transport demand, which is significantly lower than the historical car fleet. }
\label{fig:costs}
\end{figure}
Different carbon budgets do not affect the transformation speed for land transport, although the total system cost distribution for the two different carbon budgets differs, see Fig. \ref{fig:costs}. The total system cost provides an overview of the technologies chosen as cost-effective for the respective planning horizons. The annualized land transport capital costs (at the top of the graphic) show the same transition for both carbon budgets. The lower carbon budget has a higher total cost due to the more stringent carbon budget which forces the use of more wind and solar PV as well as electrolysis from 2030 on, as discussed in detail in \cite{Victoria2022}. In contrast, the \Ctwo carbon budget allows higher emissions until 2050, when the carbon limit is set to zero to ensure that both transition paths attain carbon neutrality by mid-century. Consequently, the total costs in 2050 increase significantly to install the required capacities of technologies such as electrolysis or \co capture and storage to reach carbon neutrality.\\
Disallowing DSM or V2G does not change the speed of transformation of the land transport or the share of EVs and FCEVs compared to the base scenario in which both options are active. However, compared to the base scenario where both options are activated, the total system costs increase for the deactivation of V2G by 0.1\% and for the deactivation of DSM by 1.6\%. Without DSM the land transport demand has to be provided exactly when it is demanded (we assume that this happens right after driving) without allowing the car battery to be used to store electricity when electricity prices are low. When DSM and V2G are not available, additionally flexibility must be provided, leading to an increase in stationary battery by 61\% (reaching 3008 GWh) and 9\% (reaching 2040 GWh), respectively, for 2050 compared to the Base scenario (1871 GWh). Assuming that 100\% of the EVs participate in DSM (the base scenario assumed only 50\%) reduces the total system costs by an additional 0.7\% and reduces the capacity of stationary batteries down to 1245 GWh in 2050.\\
\begin{figure}[htbp]
             \includegraphics[width=\textwidth]{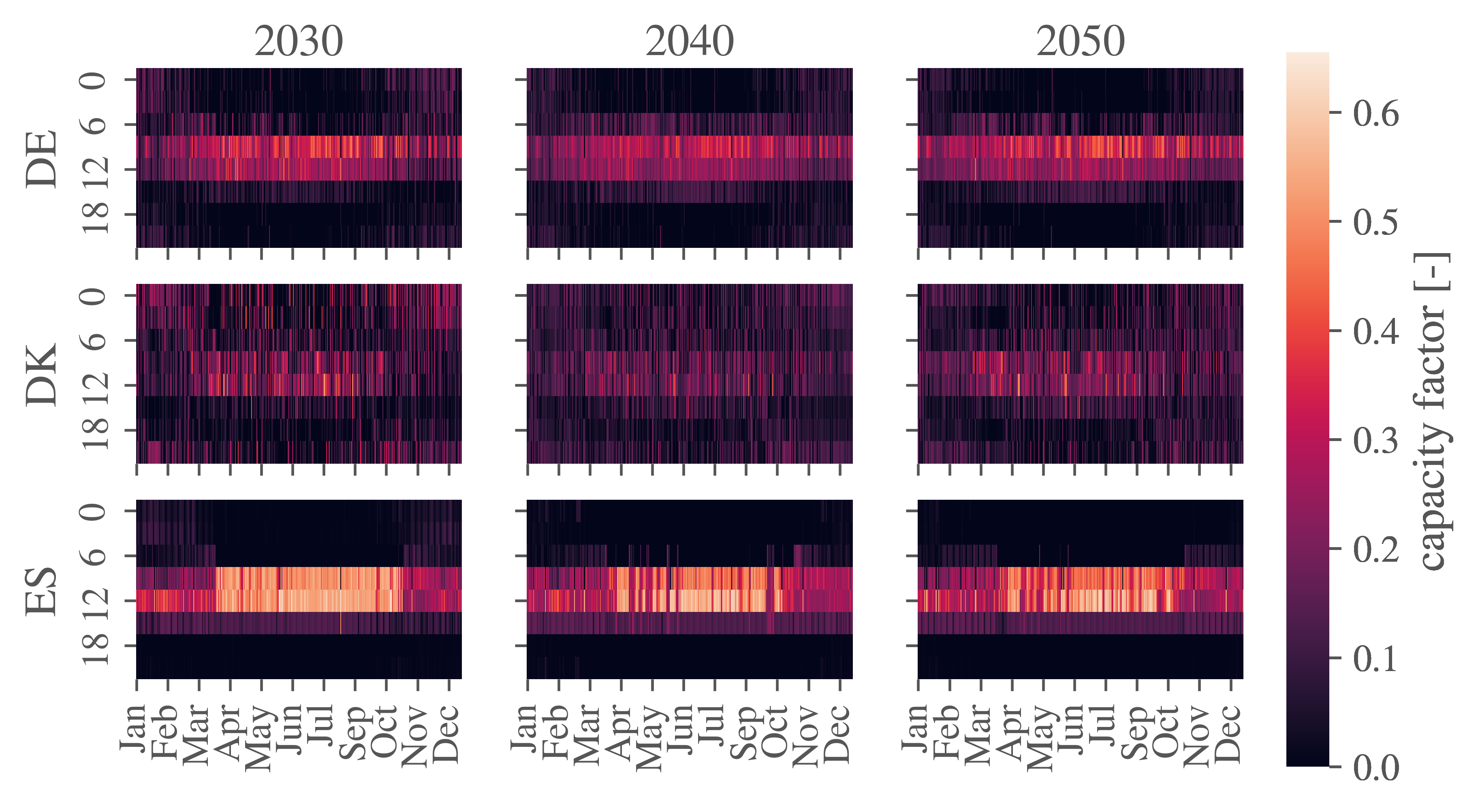}
\caption{\textbf{EV charging patterns:} EVs' daily use of charging capacity relative to installed capacity during a year for the \Cone transition path for (top) Germany, (middle) Denmark, and (bottom) Spain. See Fig. \ref{fig:v2gdaily} for an equivalent display of discharge patterns. }
\label{fig:BEVdaily_DSM}
\end{figure}
EVs are balancing the grid through smart charging and V2G use. We compare the daily and seasonal patterns of EV charging in Germany, Denmark, and Spain while allowing DSM and V2G, see Fig. \ref{fig:BEVdaily_DSM}. EV charging in Spain follows a solar-dominated charging pattern that becomes stronger as the transition progresses, while EV charging in Denmark shows a strongly wind-dominated pattern, where charging patterns depend on synoptic fluctuations in wind generation. Germany shows a combination of both patterns. These patterns confirm that DSM allows to charge the EVs in times when there is a high renewable feed-in and electricity prices are low.
 \begin{figure}[H]
\centering
    \begin{subfigure}[b]{\textwidth}
             \centering      
             \includegraphics[clip, trim=0cm 0.3cm 0cm 0cm,width=\textwidth]{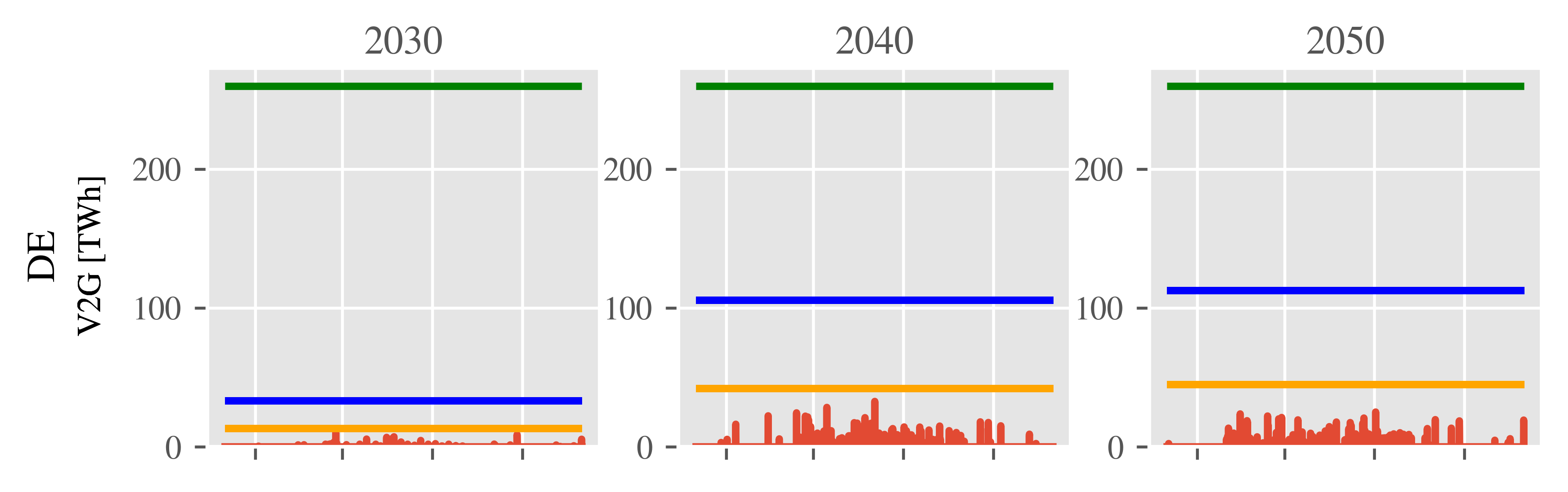}
           %  \caption{}
             \label{fig:sde25750v}
         \end{subfigure}
         \begin{subfigure}[b]{\textwidth}
             \centering
             \includegraphics[clip, trim=0cm 0.1cm 0cm 0cm,width=\textwidth]{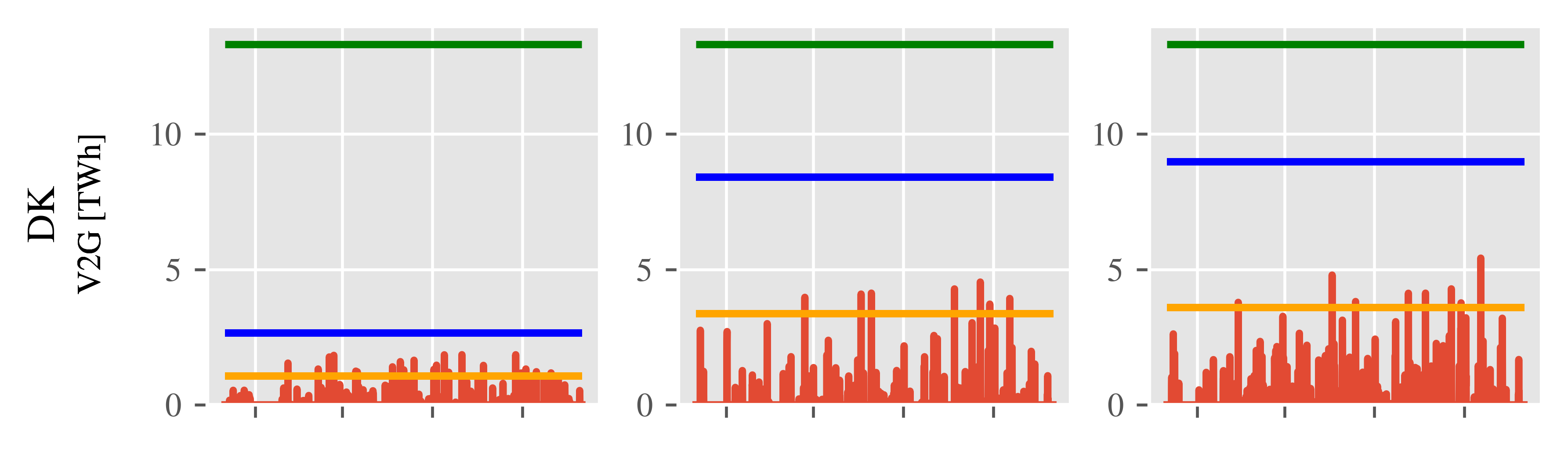}
            % \caption{}
             \label{fig:sdk25750v}
             \end{subfigure}
            \begin{subfigure}[b]{\textwidth}
             \centering
             \includegraphics[clip, trim=0cm 0cm 0cm 0cm,width=\textwidth]{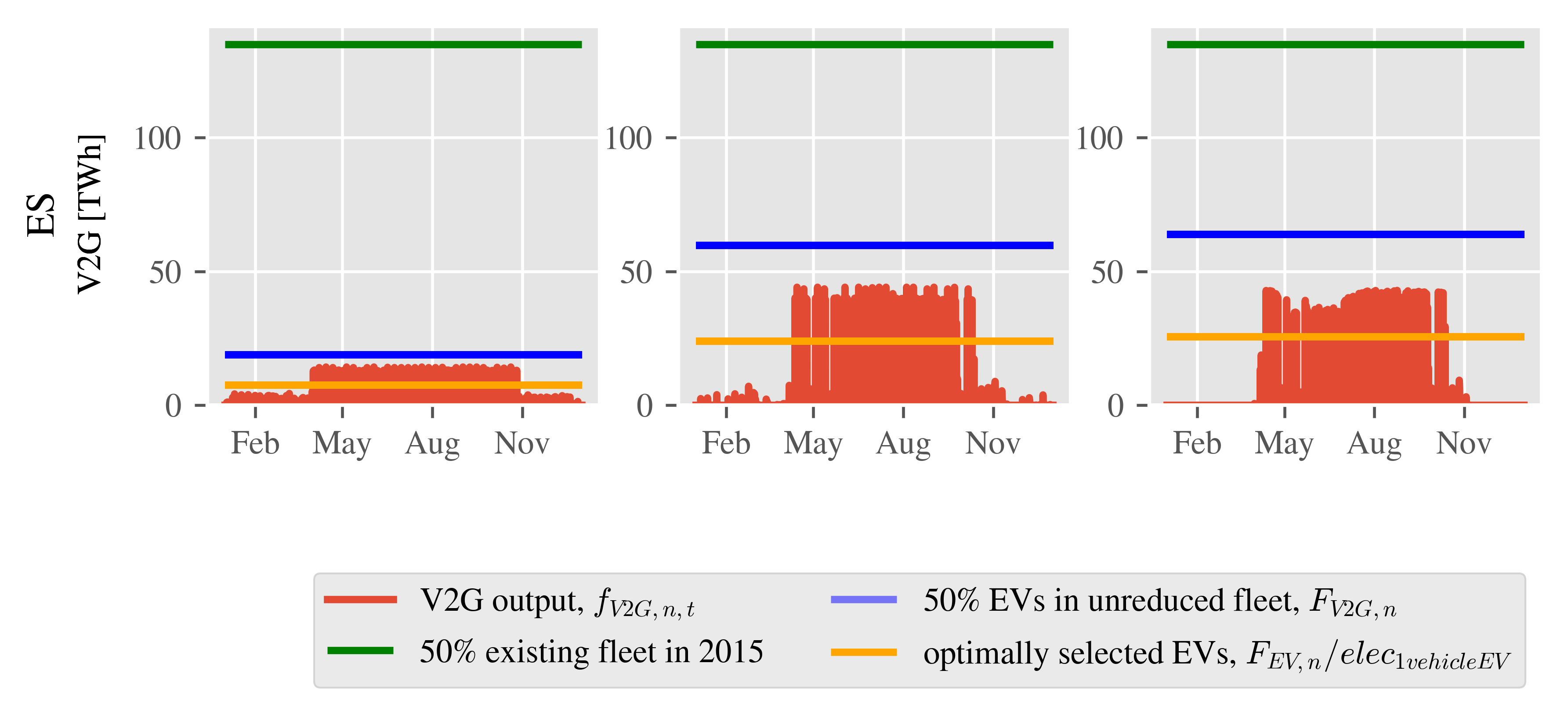}
             %\caption{}
             \label{fig:ses25750v}
         \end{subfigure}
\caption{\textbf{EVs discharging patterns:} Electricity discharged by EV batteries into the grid for a \protect\Cone transition path for (top) Germany, (middle) Denmark, and (bottom) Spain for 2030, 2040, and 2050. 
For context, three horizontal lines are drawn: The orange one indicates the V2G capacity of the EVs cost-optimally selected (\protect$F_{EV,n}/elec_{1vehicleEV}\protect$ in Eq.~\protect\ref{eq:BEV_charger}). The blue line indicates the V2G capacity of the more inefficient and realistic number of EVs (\protect$F_{V2G,n}\protect$ in Eq.~\protect\ref{eq:BEV_charger} and \protect\ref{eq:dsm}). The green line represents the V2G capacity assuming that half of the existing car fleet in 2015 was converted into EVs in every country \cite{JRC-IDEES}.}
\label{fig:v2g_DSM}
\end{figure}
We compare how Germany, Denmark and Spain use the V2G option in the base scenario, see Fig. \ref{fig:v2g_DSM}. In 2030, all countries use V2G, especially during the summer months. Similarly to charging, discharging patterns are mostly affected by solar fluctuations in Spain, wind in Denmark and a combination in Germany. For Denmark and Spain, the maximum power delivered by V2G is higher than the lumped power capacity of EVs optimally selected ($F_{V2G,n}$ in Eq. \ref{eq:BEV_charger} and \ref{eq:dsm}) but notably lower than the lumped power capacity of the unreduced and more realistic number of EVs. This means that, from a system point of view, more EVs are used for balancing than those strictly necessary to supply the transport demand. However, this balancing can be achieved using less than what 50\% participation of an unreduced electrified fleet could provide.\\
\subsection{Sensitivity to cost and fuel consumption assumptions.}
\label{sec:2}
\begin{figure}[H]
\centering
 \begin{subfigure}[b]{\textwidth}
             \centering
             \caption{}
             \includegraphics[width=\textwidth]{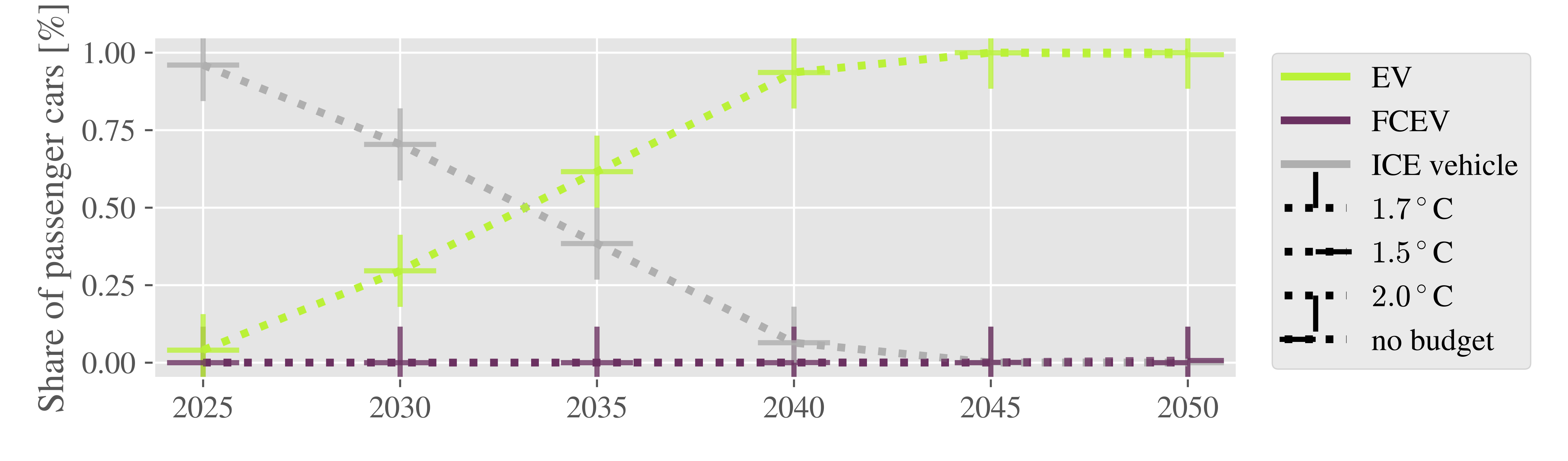}
             \label{fig:car15}
         \end{subfigure}
    \begin{subfigure}[b]{\textwidth}
            \caption{}
            \includegraphics[width=1.00\textwidth]{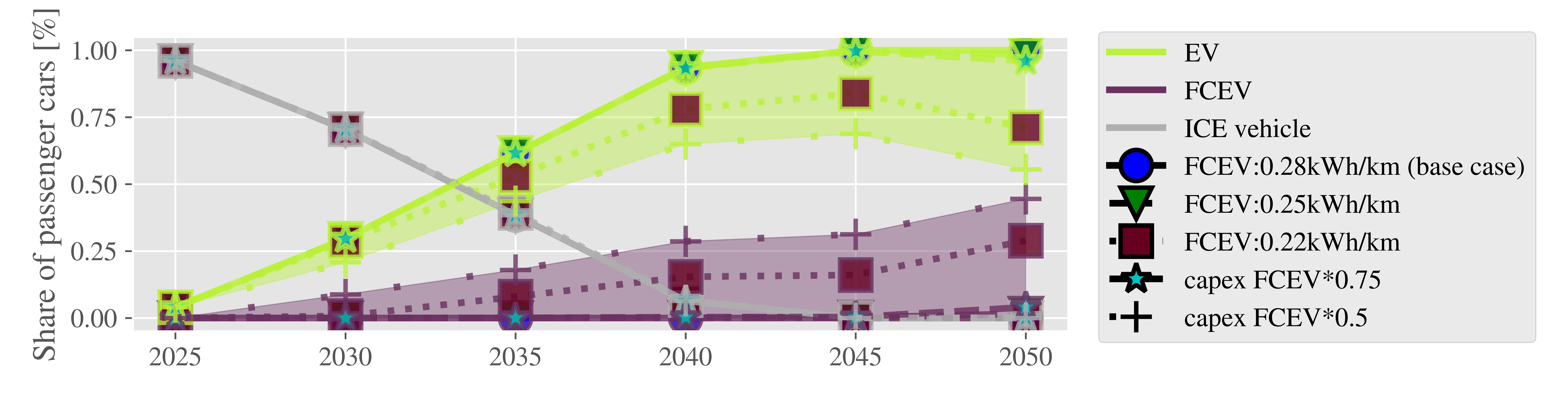}
            \label{fig:bevavail}
\end{subfigure}
\caption{\textbf{Influence of carbon budgets and FCEV capital costs and consumption}: (\subref{fig:car15}): Comparison of the base scenario to variations in the carbon budgets (\subref{fig:bevavail}): Comparison of the base scenario to different FCEV capital costs and cases where the difference in consumption between FCEVs and EVs is reduced (consumption defined in Table \ref{tab:parameters}).}
\label{fig:price}
\end{figure}
The results are robust against changes in the carbon budget, see Fig. \ref{fig:car15}. Even without a carbon budget, the transition to EVs is cost-effective. To further understand the model's preference for EVs over FCEVs, we explore the influence of EV and FCEV capital costs. A reduction of FCEV capital costs by 25\% or an increase of 25\% of capital costs for EVs does not change the share ratio, see Fig. \ref{fig:bevavail} and Fig. \ref{fig:supcapcost}. Only for large variations of the capital costs by 50\% are around 45\% of the total fleet in 2050 FCEVs. The transition is also indifferent to a capital cost increase by 50\% for both EVs and FCEVs or the increase of EV capital costs by 25\%, see \ref{Sup:4}. This is not only because the difference in capital cost of EV and FCEV is notable, but also because EVs provide balancing to the system, decreasing the need for stationary batteries and reducing the system cost, as previously discussed. Another possible reason are the operational cost differences of FCEVs and EVs due to fuel costs and efficiency. To understand the influence of the fuel use efficiency and with that indirectly the influence of fuel costs, we varied the efficiency of FCEVs, see Fig. \ref{fig:bevavail} and Fig. \ref{fig:supcons}. We assume that EVs are always more efficient than FCEVs since they do not have to convert hydrogen to electricity. Even if the efficiency gap is notably reduced, so that the FCEV consumes only 10\% more kWh of hydrogen per km, relative to the electricity consumption of EVs, less than 34\% of the total fleet are FCEVs in 2050.\\
\subsection{Regional differences}
\label{sec:3}

\begin{figure}[H]
 \begin{minipage}{\textwidth}
 \hspace{0.5in} 2025 \hspace{1.1in} 2035 \hspace{1.05in} 2050
 \end{minipage}
\raisebox{-3ex}{
\rotatebox{90}{
\begin{minipage}[b][0.03\textheight][b]{0.19\textwidth}
        \small{base scenario}
\end{minipage}
}}
\begin{minipage}{0.96\textwidth}
    \centering
         \begin{subfigure}[b]{\textwidth}
             %\centering
             \caption*{}
             \includegraphics[width=\textwidth]{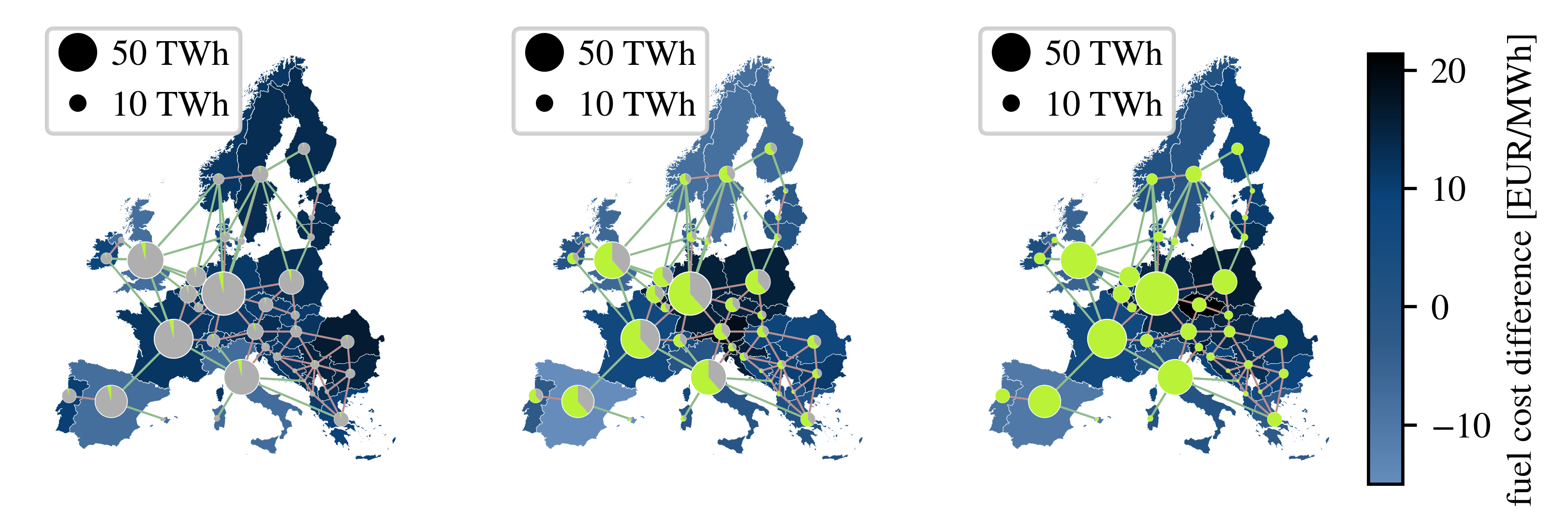}
             %\caption{}
             \label{fig:25750}
         \end{subfigure}
\end{minipage}
\rotatebox{90}{
\begin{minipage}[b][0.03\textheight][b]{0.19\textwidth}
        \small{FCEV capex*0.5}
\end{minipage}}
\begin{minipage}{0.96\textwidth}
\centering
        \begin{subfigure}[b]{\textwidth}
             \centering
             \includegraphics[width=\textwidth]{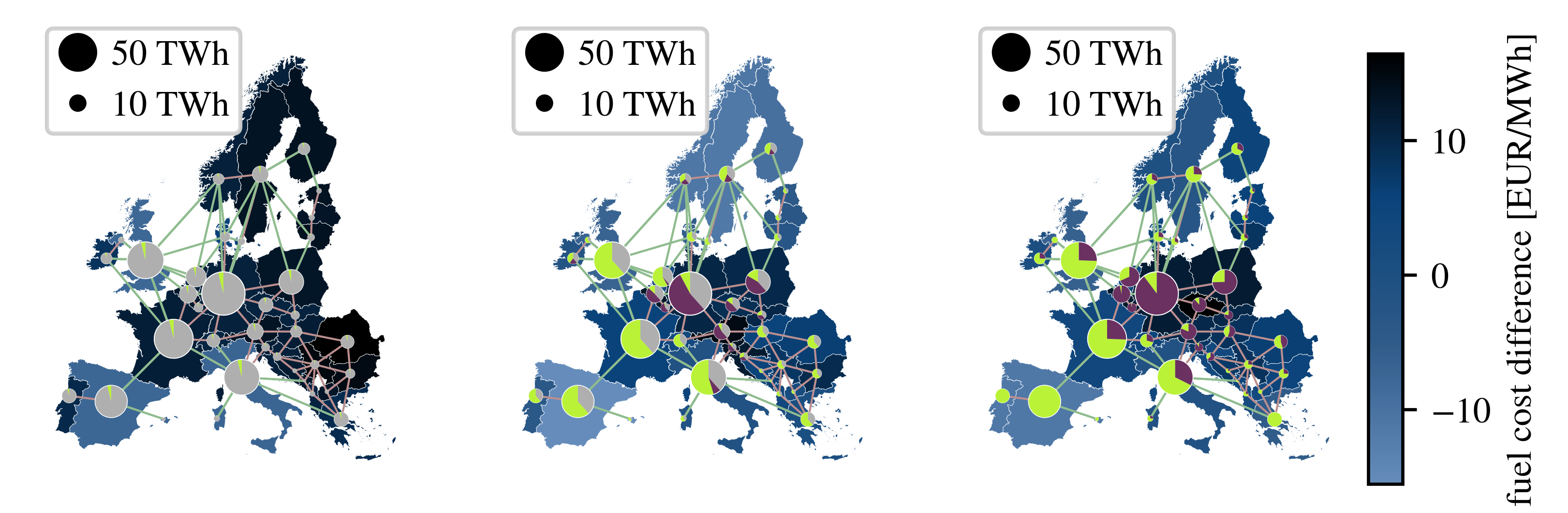}
             %\caption{}
             \label{fig:73950}
         \end{subfigure}
\end{minipage}
        \caption{\textbf{Regional distribution of vehicle ratios:} Transitions are shown from left to right for the years 2025, 2035, and 2050. The upper row shows the transition per country for the base scenario. The lower row shows the transition per country for a FCEV capital cost decrease by 50\%. The pie charts show the land transport demand in TWh, and the vehicles used to provide that demand per node, with ICE vehicles in gray, EVs in green, and FCEVs in violet. The countries' colors show the difference of the electricity and hydrogen prices, weighted by the land transport demand. Positive numbers indicate higher electricity prices. The lines between nodes show the main transmission lines.\\
The fleet was set exogenously to 96\% ICE vehicles and 4\% EVs for 2025 (see maps on the left).}
        \label{fig:spatial}
\end{figure}
In 2035, retired ICE vehicles are fully replaced by EVs for the base scenario. In contrast to passenger cars, FCEV capital costs for freight transport are predicted to be lower than EVs \cite{Sterchele2020}. If capital costs for FCEVs in freight transport become around 25\% lower, the ratio is similar to a 50\% capital cost reduction for passenger cars. In this case, FCEVs are cost-optimally selected and their use in Europe shows regional differences, see Fig. \ref{fig:spatial}. In contrast, for the 50\% FCEV capital cost reduction case FCEVs are installed in Central Europe, notably Germany, Austria and Poland. In both scenarios, countries with high solar capacities like Spain, Portugal, or Greece prefer an EV fleet. A possible explanation is the non-uniform distribution of electricity and hydrogen prices across Europe. In 2035 and 2050, electricity in Central Europe is more expensive than hydrogen during times of high land transport demand. 

\section{Discussion}
The model favors a fast electrification independent of the assumptions on EV capital costs, carbon budgets, the activation of V2G, or DSM availability. The results are not only robust independently of the carbon budgets but even without setting a carbon budget, highlighting the cost-effectiveness of EVs. Overall EVs are preferred in our base scenario. Lower capital costs and higher efficiencies in combination with direct balancing of renewables make the choice of EVs likely more attractive than FCEVs. However, FCEVs could also provide indirect balancing, when hydrogen is produced inexpensible and then stored for later use. This explains why FCEVs are used to a certain degree in some countries if FCEVs' capital costs are 50\% lower or hydrogen consumption reduces by 25\%. This could also play a role in freight transport, where capital costs of EVs and FCEVs are in a similar range \cite{DEAfreight}.\\
In line with Rinaldi et al. \cite{Rinaldi2023} who studied the Swiss transport transformation, we find electrification in the whole European system a cost-optimal strategy to decarbonize. We see a strong reduction of electricity storage when allowing DSM and V2G, similar to Bogdanov et al. \cite{Bogdanov2024} results when allowing DSM and V2G. We assume a flexibly used electrolyzer, which as Bogdanov et al. show also reduces the need for electricity storage in 2050. Our results agree also with the cost-optimal result in Fig. S15 from Pickering et al. \cite{Pickering2022}, where land transport is electrified almost completely, and smart charging is preferred.\\

ICEs are not cost-competitive even in the near-term. This stands in stark contrast to today's land transport sector in Europe. While Norway, as the leading country in electrification of the road transport, shows good agreement with the model's optimal transformation path speed, other European countries lag behind. Especially some larger European countries with lower electric car shares, notably Germany with around 2.8\% and Spain with 0.6\% in 2023 \cite{afo} will be challenged to undergo a fast transition. In contrast, Scandinavian countries like Denmark or Sweden with a share of around 7\% of EVs could follow the transition path if they keep their current transformation speed. Europe's "Fit for 55" plan allows only zero-emission passenger cars and light commercial vehicles to be sold by 2035. However, while an earlier adoption of EVs is more cost-effective from a system point of view and helps with current climate goals, consumers are hesitant due to high upfront retail prices, charging possibilities and driving range \cite{consumer}. Following Europe's "Fit for 55" plan \cite{fit455} is consequently not enough. Banning the sale of \co-emitting cars by 2035 falls short, since many ICE cars will remain in operation until the end of their lifetime. In contrast, our model has decarbonized the land transport sector by around 93\% by 2035. Slower adoption rates of EVs would put pressure on other sectors to decarbonize early and lead to the need of more expensive carbon-capture technologies. Slower adoption rates could also lead to problems in the car industry, where companies need sales security for long-term investments in EVs.\\

The main limitations of our study are presented in this paragraph. Our study does not include temperature dependent efficiencies of vehicles or infrastructure costs. Brown et al. \cite{Brown2018} included a temperature-dependent efficiency for EVs. The European Automobile Manufacturers Association (ACEA) expects infrastructure costs of €1000 billion until 2050 including expenses for transmission grid expansion and renewable installation, which would be around €51bn annually for charging points. Comparing this value to the number of existing vehicles and using the EV capital cost assumption of this paper, this increases individual EV capital costs by 9\%, see \ref{Sup:2}). The infrastructure costs are indirectly included in our cost sensitivity study. Given that an increase of up to 25\% of EV capital costs does not change the ratios between EVs and FCEVs, adding the cost of chargers and grid distribution reinforcement to EVs is also not expected to change this ratio. We do not distinguish between passenger vehicles, light commercial, heavy commercial vehicles, and 2-wheelers which have different capital costs, fuel use, and different usage profiles. While this is a limitation, we assume that vehicles are mostly used during the day and our general land transport demand profile captures the peak demands and balancing needs. FCEV capital costs for freight transport are predicted to be cheaper than EVs \cite{Sterchele2020}. If capital costs for FCEVs in freight transport become around 25\% cheaper, the ratio would be similar to a 50\% capital cost reduction for passenger cars. According to our results in Fig. \ref{fig:car15}, this could mean that FCEVs would optimally cover one fifth of 2050's freight transport demand.\\
For simplicity, we assume the same amount of land transport demand and vehicle usage profile throughout the years. However, several trends could influence this: Car sharing or increased attractiveness of public transport may decrease the land transport demand of vehicles or change the usage profiles. In contrast, historical data show an increase in land transport demand \cite{JRC-IDEES}. It is difficult to predict, which effect will be dominant. We assume one type of vehicle per fuel type and do not include any increase in vehicle sizes as observed in historical data \cite{IEA2022} which increases the fuel use per vehicle. We further assume a fixed lifetime, neglecting regional or socio-economic differences in vehicle usage. It is difficult to estimate which of these partly opposite trends will affect the land transport demand the most or if they will cancel each other. We do not consider plug-in electric vehicles, since these will not be carbon neutral and because their average capital costs are around 20\% higher compared to BEVs in Europe \cite{IEA2022}. We made no distinction between petrol and diesel ICE vehicles. The results are based on a spatial resolution with 37 nodes and 3-hourly resolution which can underestimate regional balancing needs and costs of fuel distribution. Previous analyses have shown that 3-hours is a good compromise to capture the most important system dynamics \cite{Schlachtberger2018, Schyska2021}.\\
Our study shows that the most cost-effective transition to net-zero \co emissions prioritizes the fast electrification of the land transport sector. When comparing our results to historical data, it is important to recall that the model determines the optimal number of vehicles needed to cover the land transport demand. For example, for 2025, 3.6\% of the existing ICE vehicle fleet would be enough to cover the land transport demand, assuming that the use of each vehicle is optimized. In reality, the existing fleet is larger because passenger cars provide people with autonomy and the fleet size is therefore not optimized for its transport demand. Vehicles are estimated to be parked 92-97\% of the time \cite{parkingEVs,Bates2012}. Consequently, we only account for the capital cost of an ideal fleet of cars. We have considered the more realistic and inefficient fleet to learn about how DSM participation influences the system. While EVs are attractive enough without any DSM participation to be chosen by the system, we see a 3\% total system cost reduction when DSM is activated, mostly due to the capacity reduction of stationary batteries. Even a low participation could help to reduce the need for stationary batteries to balance renewables.\\

A massive deployment of EVs also means a massive increase in electric engine and battery materials like copper (Cu) and lithium (Li). While an ICE passenger car uses around 23kg of copper, EVs need on average 83kg \cite{Copper}. Lithium is identified as the most critical material to enable the energy transition based on renewable technologies and electrification \cite{IEA2024minerals}. According to the IEA, the available supply, including projects that are currently still in development, only covers 70\% of copper and 50\% of lithium demand in 2035. Moreover, the long-term availability of lithium should be evaluated at a global scale. To that end, the global analysis performed by Greim et al. \cite{Greim2020} is relevant. The authors conclude that there would only be enough lithium to supply the estimated need up to 2100 for stationary and EV battery capacity (estimated globally at 200 and 50 TWh respectively) assuming that: (i) the number of light vehicles on the road increases to 3 billion (a population of 10 billion people with 0.3 vehicles per capita) of which 80\% are EVs, (ii) a high global lithium productivity is achieved (around 50 Mt of lithium which includes extracting all mineral resources), (iii) the batteries of EVs have a lifetime of 8 years after which they are used as stationary batteries for another 8 years, (iv) lithium is collected and recycled at a rate of 95\%. All these conditions are technologically possible but also extremely challenging \cite{Greim2020}.\\

\section{Conclusion}
Two clear policy recommendations can be derived from our results. First, historical electrification rates are much slower than those optimally calculated from a system perspective regardless of the chosen carbon budget. To accelerate electrification, stronger policies promoting the adoption of EVs and disincentivizing new sales of ICE vehicles as of today are recommended as well as policies supporting enough infrastructure to enable the electrification of the transport sector. Second, demand-side management provided by smart charging of EVs could bring significant benefits at the system-level, substantially reducing the need for stationary batteries to balance renewable fluctuations. This calls for the development of policy incentives facilitating smart charging of EVs.

\section*{Glossary}
\begin{description}
\item[V2G] vehicle-to-grid
\item[DSM] demand-side management
\item[EV] electric vehicle
\item[ICE] internal combustion engine 
\item[FVEV] fuel cell electric vehicle
\item[LULUCF] Land Use, Land Use Change and Forestry
\item[TEN-T] Trans-European Transport Network
\end{description}

\section*{CRediT author statement}
Sina Kalweit: Conceptualization, Methodology, Software, Investigation, Writing – original draft, Writing – Review \& Editing, Visualization\\
Elisabeth Zeyen: Writing - Review \& Editing, Software\\
Marta Victoria: Conceptualization, Methodology, Software, Writing – Review \& Editing, Supervision

\section*{Acknowledgments}
S.K. is fully funded and M.V. is partially funded by the Novo Nordisk CO2 Research Center (CORC) under grant number CORC005. We sincerely thank Alberto Alamia for the fruitful discussions on this paper.

\section*{Data and code availability}
The model is implemented with the open energy modeling framework PyPSA v0.26.2 and based on the model PyPSA-Eur-Sec v0.9.0, which was adapted to this paper as can be seen in https://github.com/s8au/pypsa-eur/tree/endogenous-transport. It uses the costs and technology assumptions included in the technology-data v0.6.2. \\
Datasets and visualization scripts can be accessed from the public repository: 10.5281/zenodo.14677976

%\appendix
%\section{Supplementary material}
%\label{app1}
%Supplementary information related to this article can be found online at [to be added]

\bibliographystyle{elsarticle-num} 
\bibliography{bib}

\clearpage

\setcounter{figure}{0}
\renewcommand{\figurename}{Fig.}
\renewcommand{\thefigure}{S\arabic{figure}}
\setcounter{table}{0}
\renewcommand{\thetable}{S\arabic{table}}
\setcounter{section}{0}
\renewcommand{\thesection}{Supplementary Note \arabic{section}}
\setcounter{equation}{0}
\renewcommand{\theequation}{S\arabic{equation}}
\noindent\textbf{\large{Supplementary Information}}\\
\section{Calculation of vehicle consumption}
\label{Sup:1}
For EVs we estimate fuel consumption as \SI{0.31}{kWh/Mi}=\SI{0.2}{kWh/km} based on the 2020 Tesla Model S Standard Range model \cite{Tesla2020}. Assuming an average driving velocity of \SI{50}{km/h} this translates to
\begin{equation}
    elec_{\text{1 vehicle EV}}=0.2\frac{50}{1000}\left[\frac{\text{kWh/km}\cdot\text{km/h}}{\text{kWh/MWh}}\right]=0.01\text{MWh}_{el}\text{/h/vehicle}
\end{equation}
as the electricity consumption of one EV in one hour. Since the lumped demand in the model is given in MWh, the unitary capital costs $C_{\text{1 vehicle EV}}$ need to be converted from EUR/vehicle to EUR/MW:
The unitary capital cost for the link representing the lumped EVs is calculated as
\begin{equation}
    \tilde{C}[\text{EUR/MW}]=\frac{C_{\text{1 vehicle EV}}[\text{EUR/vehicle}]}{elec_{\text{1 vehicle EV}}[\text{MWh}_\text{el}/\text{h/vehicle]}}
\end{equation} 
where $C_{\text{1 vehicle EV}}$ is the capital cost of 1 EV from Tab.\ref{tab:parameters}.
A vehicle with higher energy consumption may appear less expensive (in units of EUR/MW) but because its efficiency is lower, it requires a higher capacity for the link representing the lumped number of vehicles to supply the transport demand.\\
For a FCEV, we base the fuel consumption on the Hyundai NEXO with a consumption of \SI{0.0084}{kg/km} \cite{Hyundai-nexo}. Using the lower heating value of hydrogen of \SI{33.33}{kWh/kg}, this amounts to 
\begin{equation}
    0.84\text{kg}_{H_2}\text{/100km}\cdot33.33\text{kWh/kg}_{H_2}=0.28\text{kWh/km}.
\end{equation}
This leads to a power use per vehicle of 
\begin{equation}
    hydrogen_{\text{1 vehicle H}_2}=0.28\frac{50}{1000}=0.014\text{MWh}_{H_2}\text{/h/vehicle}.
\end{equation}
For ICE vehicles, we assume an average vehicle fuel consumption of gasoline of \SI{0.069}{L/km} and an energy density of \SI{34.2}{MJ/L}=\SI{9.5}{kWh/L} \cite{Mazloomi2012}. This leads to $0.069\cdot 9.5=0.66\:$kWh/km. With this we can calculate the oil use per vehicle as 
\begin{equation}
    oil_{\text{1 vehicle ICE}}=0.66\frac{50}{1000}=0.033\text{MWh}_{\text{oil}}\text{/h/vehicle}.
\end{equation}

\section{Calculation of EV infrastructure costs}
\label{Sup:2}
Assuming that 40\% of the €1000bn investment will be spent on renewables and transmission, which is already included in our total system cost, this leads to an investment of €600bn over 25 years. With a discount rate of 5\%, this leads to €51bn annually investment to install public and private chargers. This is based on the existing number of vehicles, including freight transport. Assuming average capital costs for EVs of 24879€/vehicle and a existing fleet of 0.26bn vehicles, this leads to an capital cost increase of 9\% per vehicle. 

\section{Description of land transport in PyPSA-Eur}
\label{Sup:3}
In the model, we use the PyPSA component "link" to represent the lumped sum of vehicles of a certain type (Fig.\ref{fig:links}). A link can convert energy with a certain efficiency. For instance, the "EVs" link can transform electricity into kinetic energy demanded in the land transport sector. Three links named "EVs", "FCEVs" and "ICE vehicles" exist in every region. Every link transforms the energy source (electricity, hydrogen or gasoline/diesel) into kinetic energy, with an efficiency that is inversely proportional to the vehicle consumption described in Table 1. Moreover, in PyPSA-Eur, the unitary capital cost of every link is expressed in EUR/MW, that is, it represents the cost of the exact number of cars that are required to transform 1 MWh of electricity into kinetic energy for the EVs, and similarly for FCEVs and ICE vehicles. To calculate the unitary capital cost of every link, we can divide the cost of one EV by the electricity consumption of the EV in one hour. The later is estimated as 0.01 MWh, assuming an EV consumption of 0.2 kWh\textsubscript{el}/km and average driving speed of 50 km/h, see \ref{Sup:1} for equivalent calculations for FCEVs and ICE vehicles.\\
The capacity of every link, that is, the number of vehicles of a certain type, is optimized. Several constraints were already introduced in previous publications, which we describe here for the readers convenience.
It is ensured that the land transport demand $d_{n,t}$ for every node $n$ and time step $t$ is equal to the sum of the transport demand provided by ICE vehicles $f_{\text{ICE},n,t}$, FCEVs $f_{\text{H\textsubscript{2}},n,t}$ and EVs $f_{\text{EV},n,t}$
\begin{equation}
    f_{\text{ICE},n,t} + f_{\text{EV},n,t} + f_{H\textsubscript{2},n,t} = d_{n,t}\quad \forall n,t \label{eq:demand}.
\end{equation}
DSM basically decouples charging and using of the EV battery in time. For this time-decoupling, an EV battery is added to the model, as shown in Fig. \ref{fig:links}. The EV battery decouples the use of the link representing the "EV battery charger" and the link "EVs" that transform electricity into kinetic energy in time. If DSM is activated, the state of charge of the EV batteries $e_{EV battery,n}$ at 6am needs to be at least 75\% of their available capacity $E_{EV battery,n}$ (Eq. \ref{eq:min_dsm}) to represent vehicles should be ready to be used for commuting. This avoids that EV batteries are used as seasonal storage.
 \begin{equation}
     e_{\text{EV battery},6:00,n} \geq 0.75\:E_{\text{EV},n}\quad \forall n\label{eq:min_dsm}
 \end{equation}
For V2G use in the model, an additional link is added that allows discharging electricity from the EV battery to the power grid. To represent the behavior of the lumped EVs in a region, several constraints are added. First, to represent the charging of EVs, a charger with power capacity $r_{\text{charge}}=$\SI{11}{kW} is assumed for every vehicle. Second, to represent the percentage of EVs which are parked every hour, and hence can potentially be charged, we define an availability profile which is inversely proportional to the demand profile (Fig. \ref{fig:profile}). The share of available cars $av_{n,t}$ in every node $n$ and time step $t$ is calculated assuming a maximum availability $av_{\text{max}}=0.95$ and mean availability $av_{\text{mean}}=0.8$:
\begin{equation}
    av_{n,t} = av_{\text{max}} - \frac{(av_{\text{max}}-av_{\text{mean}})(d_{n,t}-min(d_{n,t}))}{mean(d_{n,t})-min(d_{n,t})}\quad\forall n,t
    \label{eq:avail}
\end{equation}

\section{Sensitivity analysis}
\label{Sup:4}

\begin{figure}[htbp]
             \includegraphics[width=\textwidth]{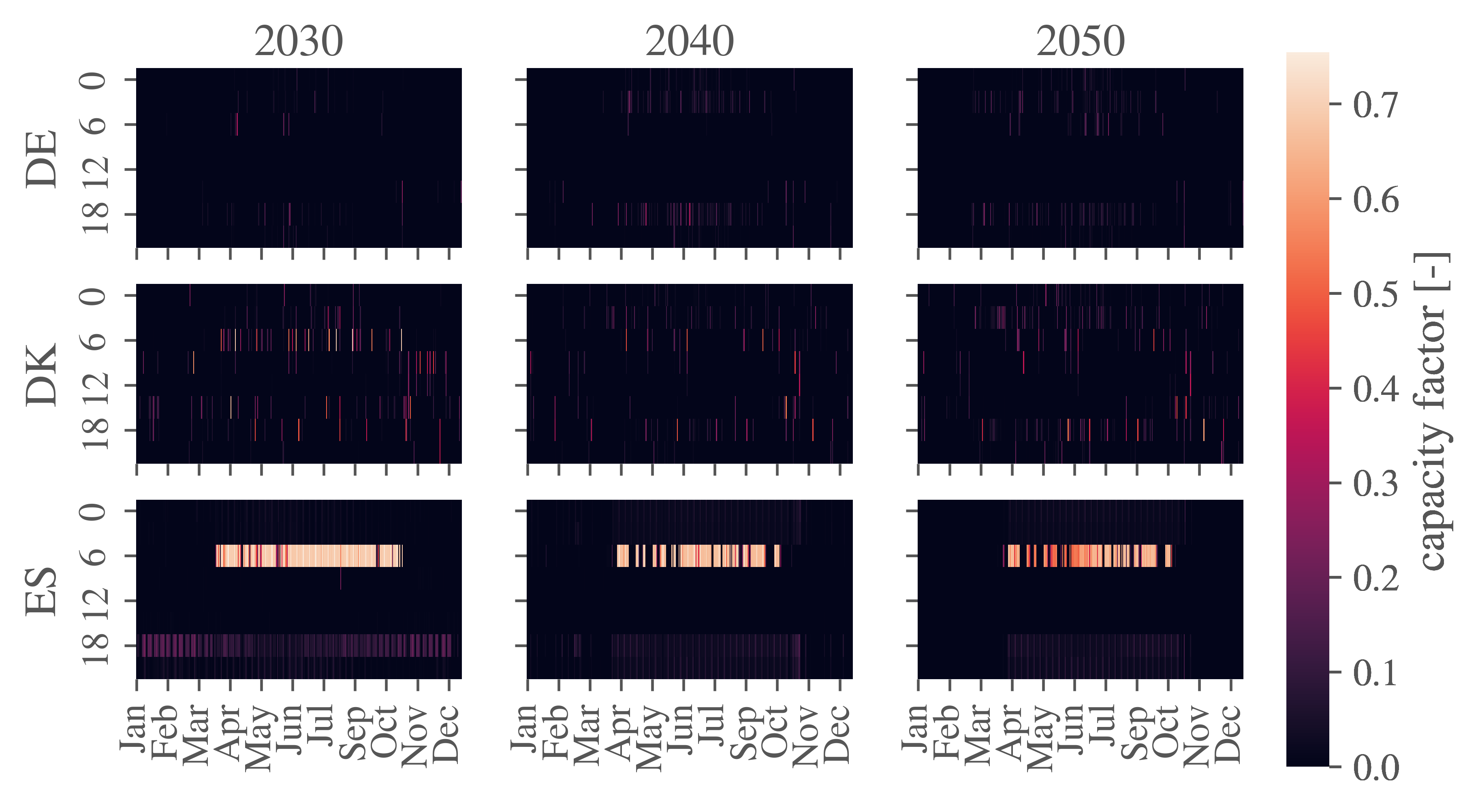}
\caption{\textbf{EV discharging patterns:} EVs' daily V2G use during a year for the \Cone transition path for (top) Germany, (middle) Denmark, and (bottom) Spain. }
\label{fig:v2gdaily}
\end{figure}

\begin{figure}
    \centering
    \includegraphics[width=\linewidth]{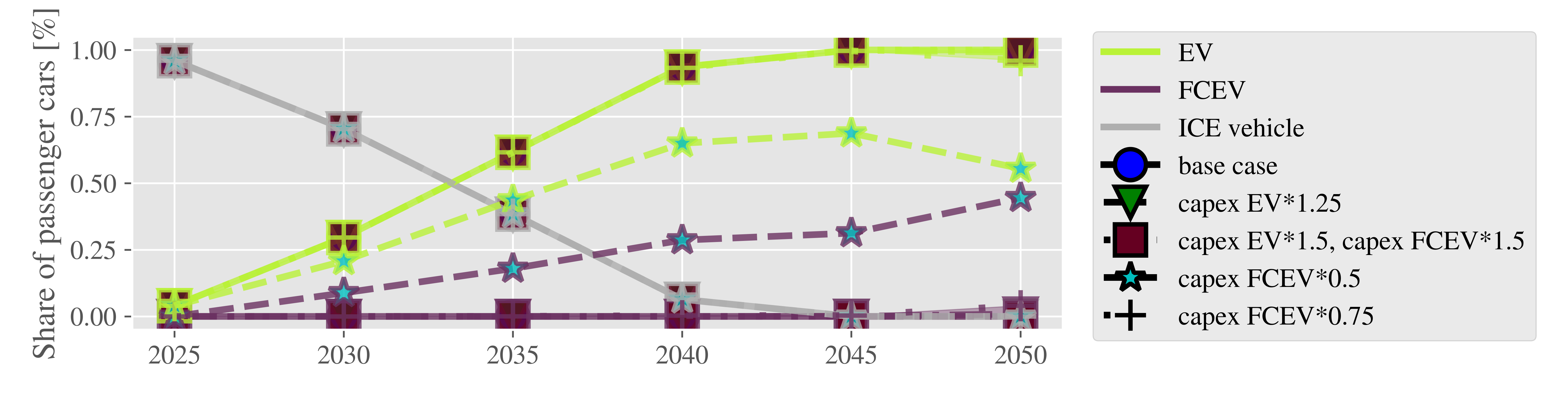}
    \caption{Transition path for the \Cone scenario with capital costs variations.}
    \label{fig:supcapcost}
\end{figure}

\begin{figure}
    \centering
    \includegraphics[width=\linewidth]{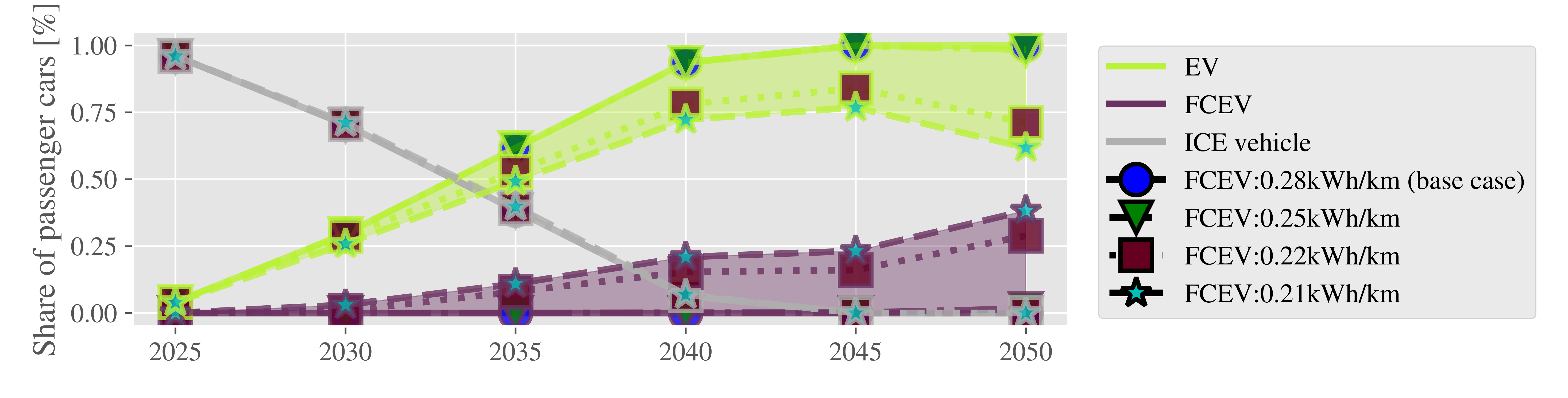}
    \caption{Transition path for the \Cone scenario with variations in FCEVs' consumption.}
    \label{fig:supcons}
\end{figure}

\begin{figure}
    \centering
    \includegraphics[width=\linewidth]{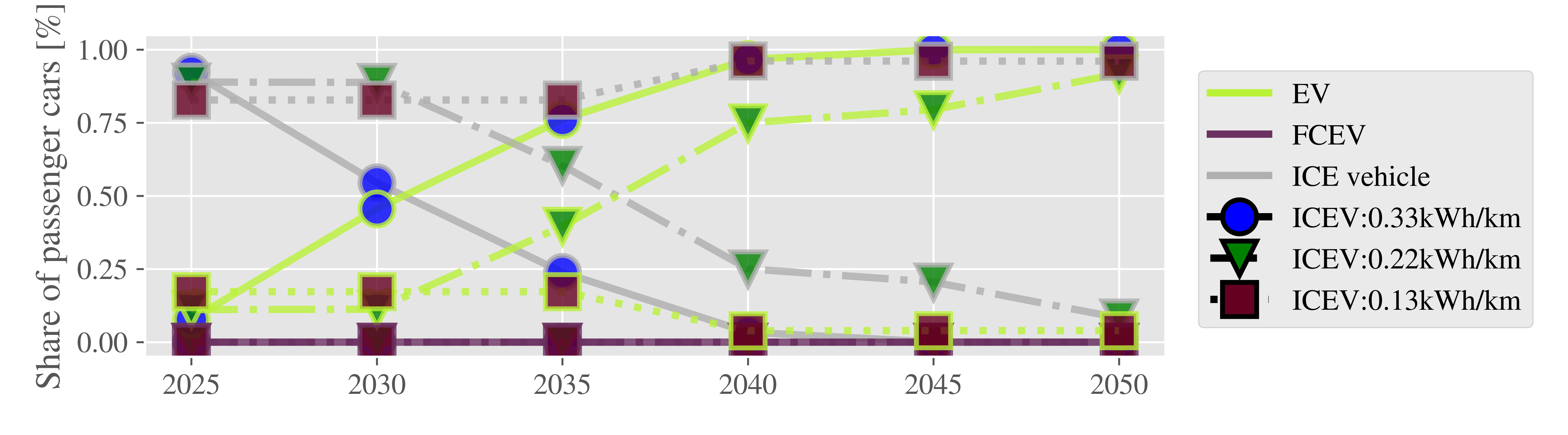}
    \caption{Transition path for the \Cone carbon budget case, where ICE vehicle consumption is varied.}
    \label{fig:ICEvehcons}
\end{figure}

\begin{figure}
    \centering
    \includegraphics[width=\linewidth]{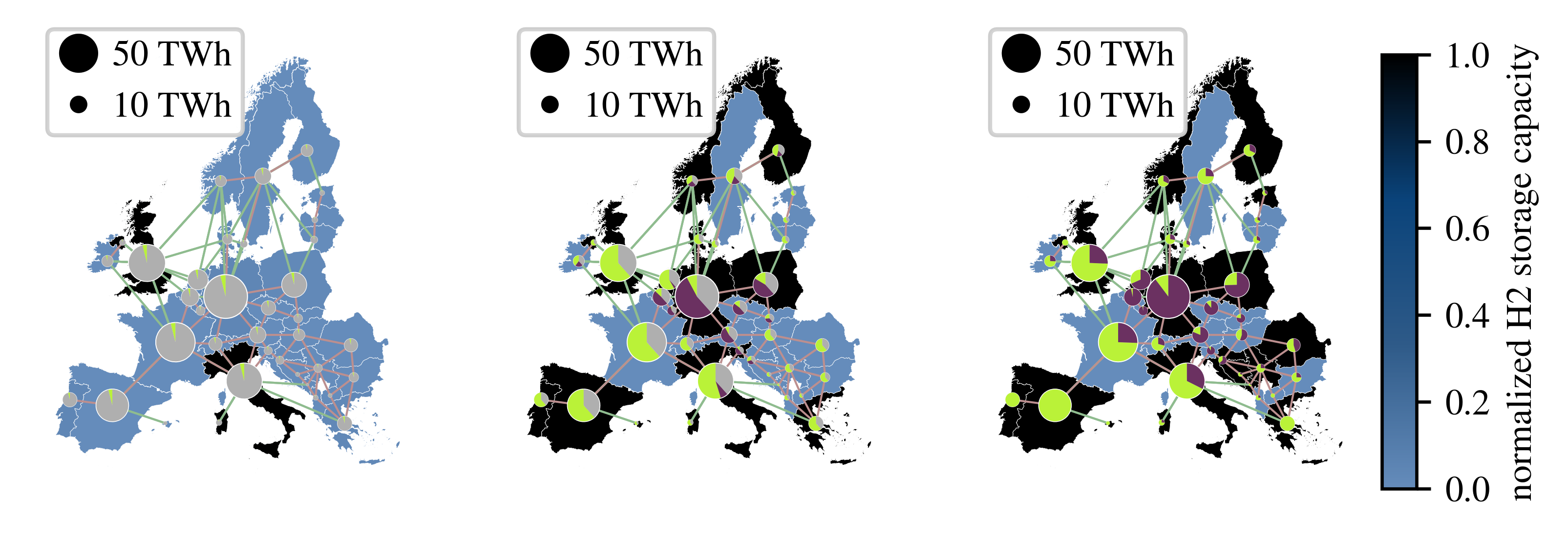}
    \caption{Transition path for the \Cone carbon budget case, where FCEV capital costs are 50\% reduced. The colors of the countries show the installed H\textsubscript{2} storage capacity.}
    \label{fig:H2storageregional}
\end{figure}

\begin{figure}
    \centering
    \includegraphics[page=3,width=\linewidth]{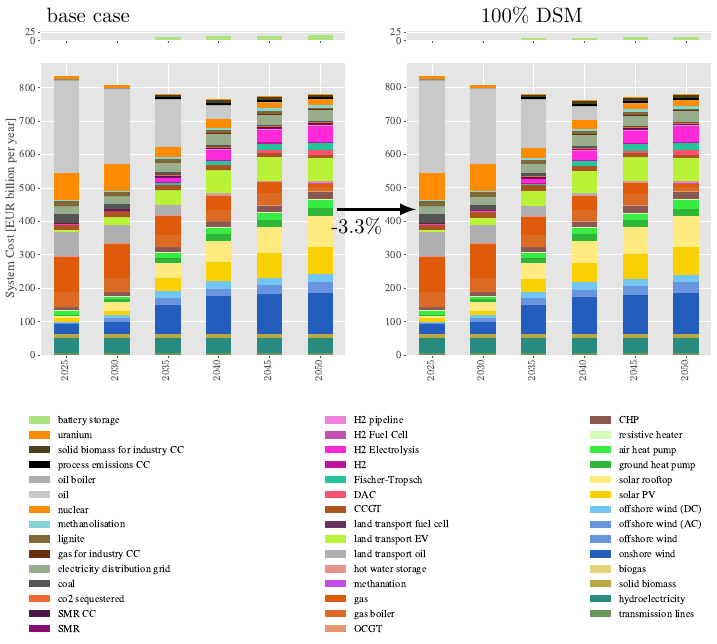}
    \caption{\textbf{Total system costs for different DSM availability:} The plots show the total system cost for the base case with 50\% DSM participation (left), a case where the whole fleet is participating in DSM (right). If more of the fleet participate in DSM, less wind and less electrolysis capacity is used on top of using fewer stationary batteries (shown separately on top). }
    \label{fig:battery_storage1}
\end{figure}
\begin{figure}
    \centering
    \includegraphics[page=1,height=0.9\textheight,keepaspectratio]{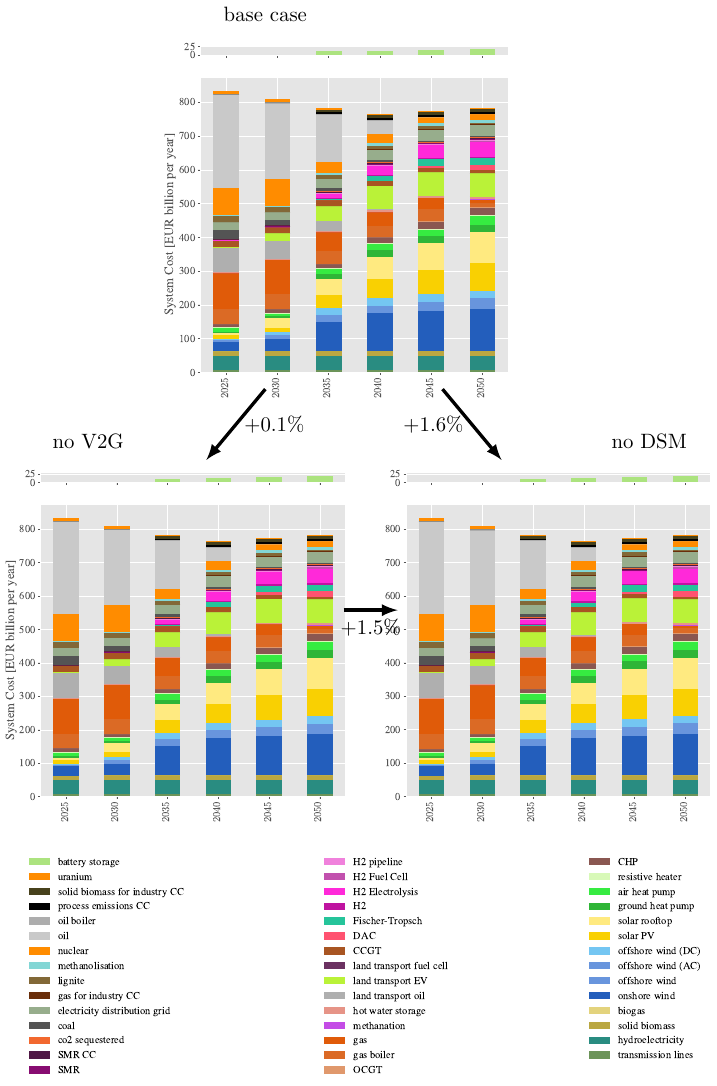}
    \caption{\textbf{Total system costs for different DSM availability:} The plots show the total system cost for the base case with 50\% DSM participation (top), a case without V2G (bottom left) and a case without DSM and V2G (bottom right). If V2G and DSM are deactivated, the most notable change is the increase in stationary battery capacity. }
    \label{fig:battery_storage2}
\end{figure}
\clearpage
{\small

%\bibliography{supplements}

\end{document}